Fig.1(a)

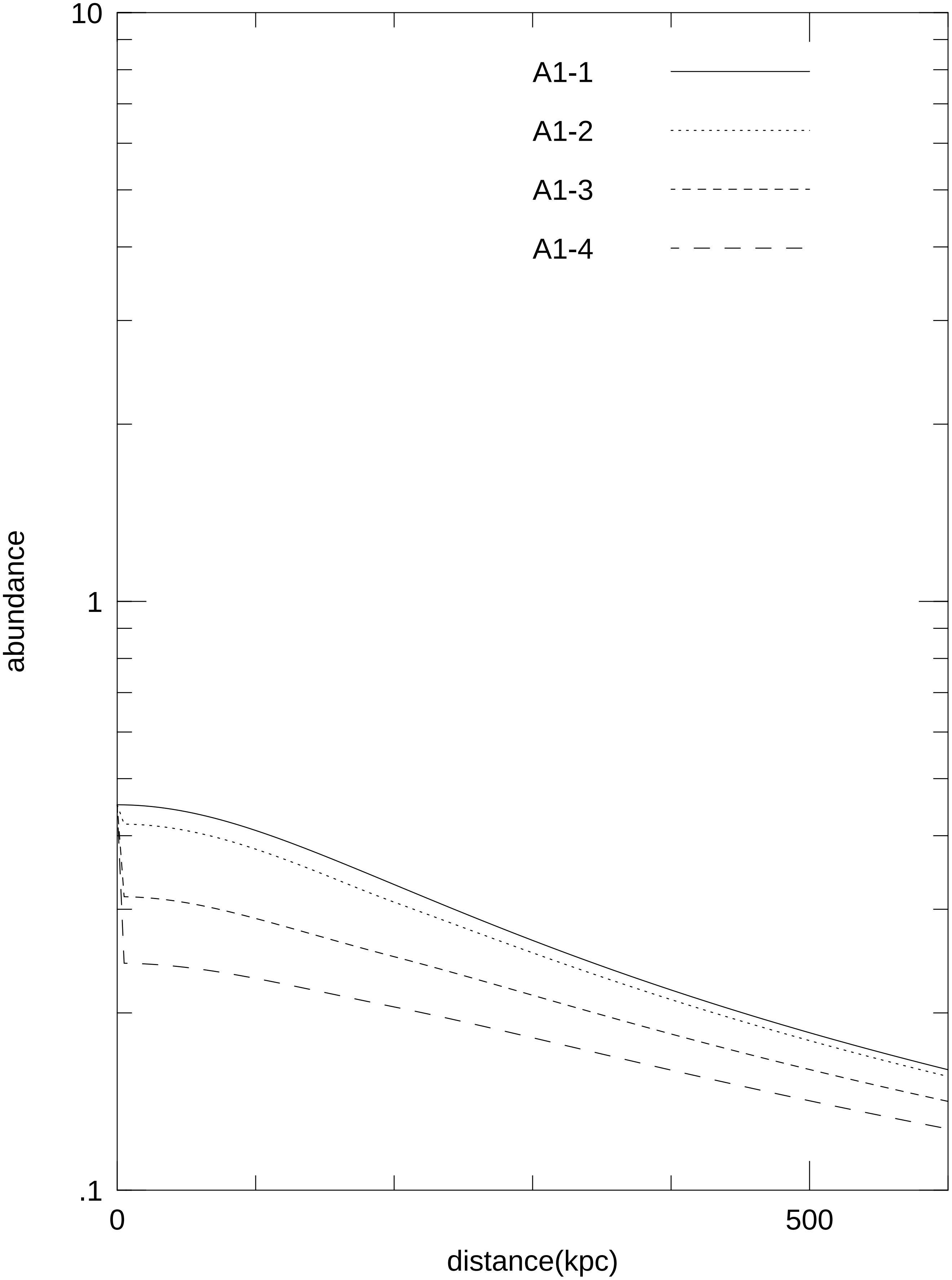

Fig.1(b)

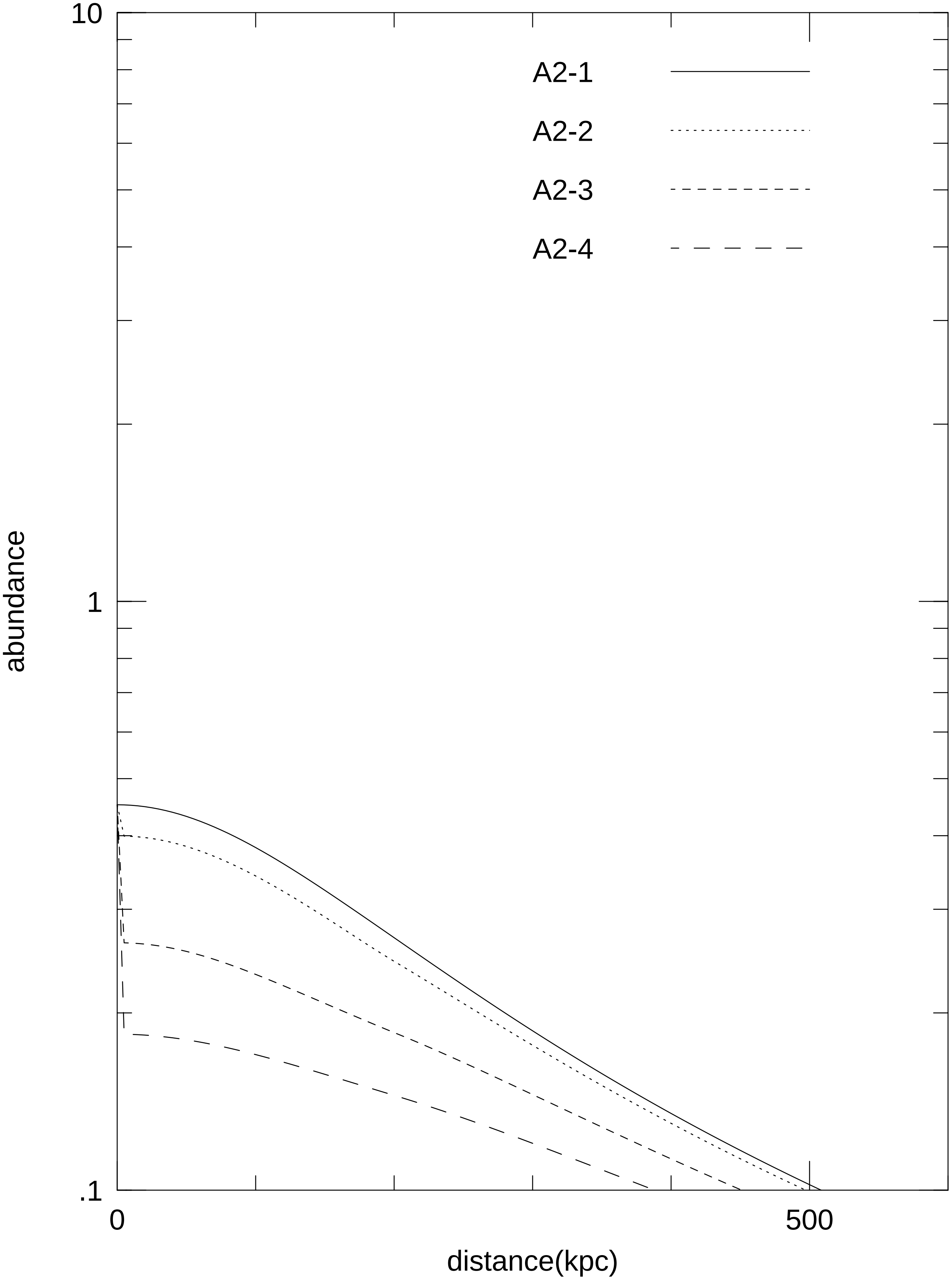

Fig.2(a)

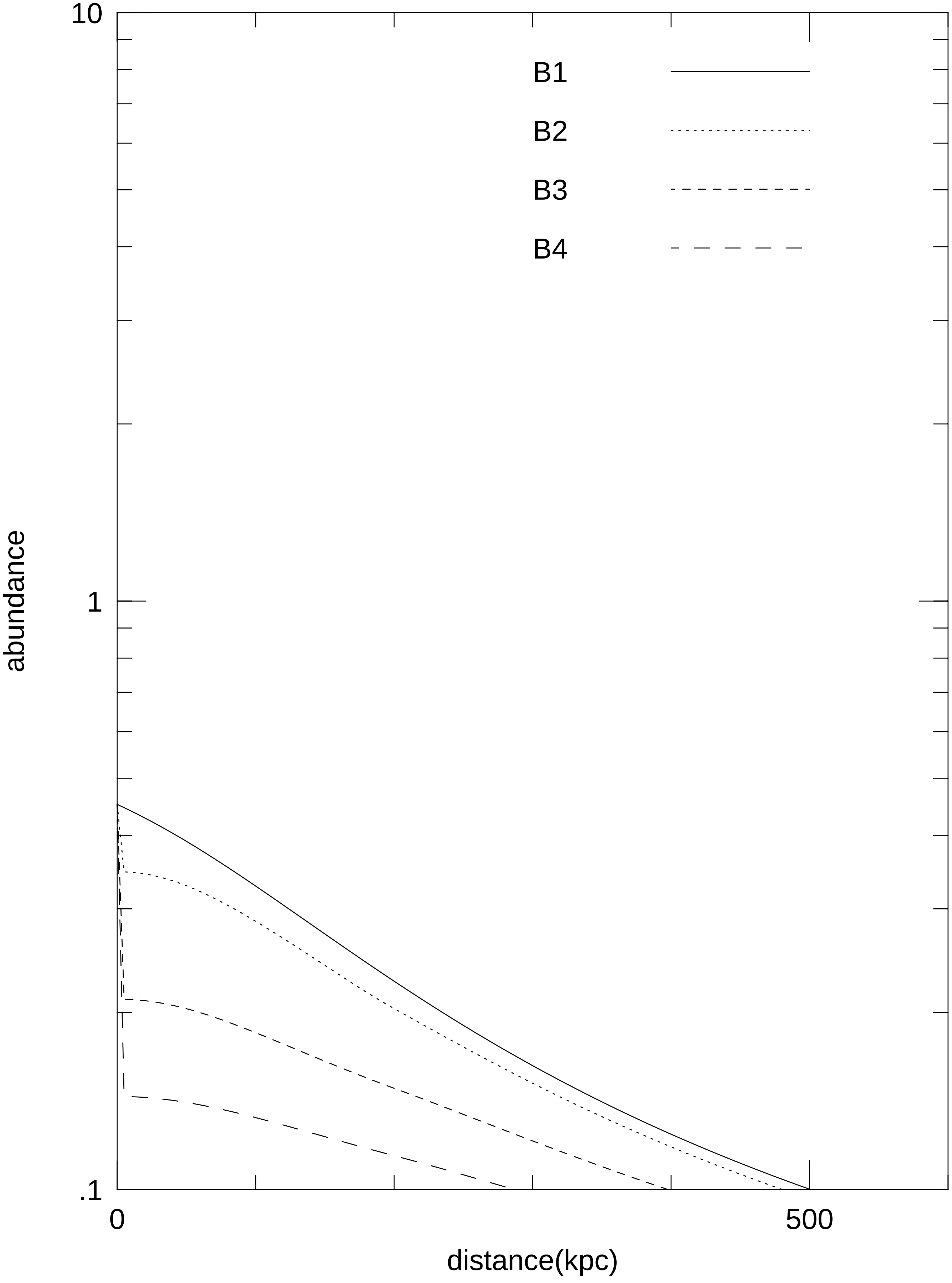

Fig.2(b)

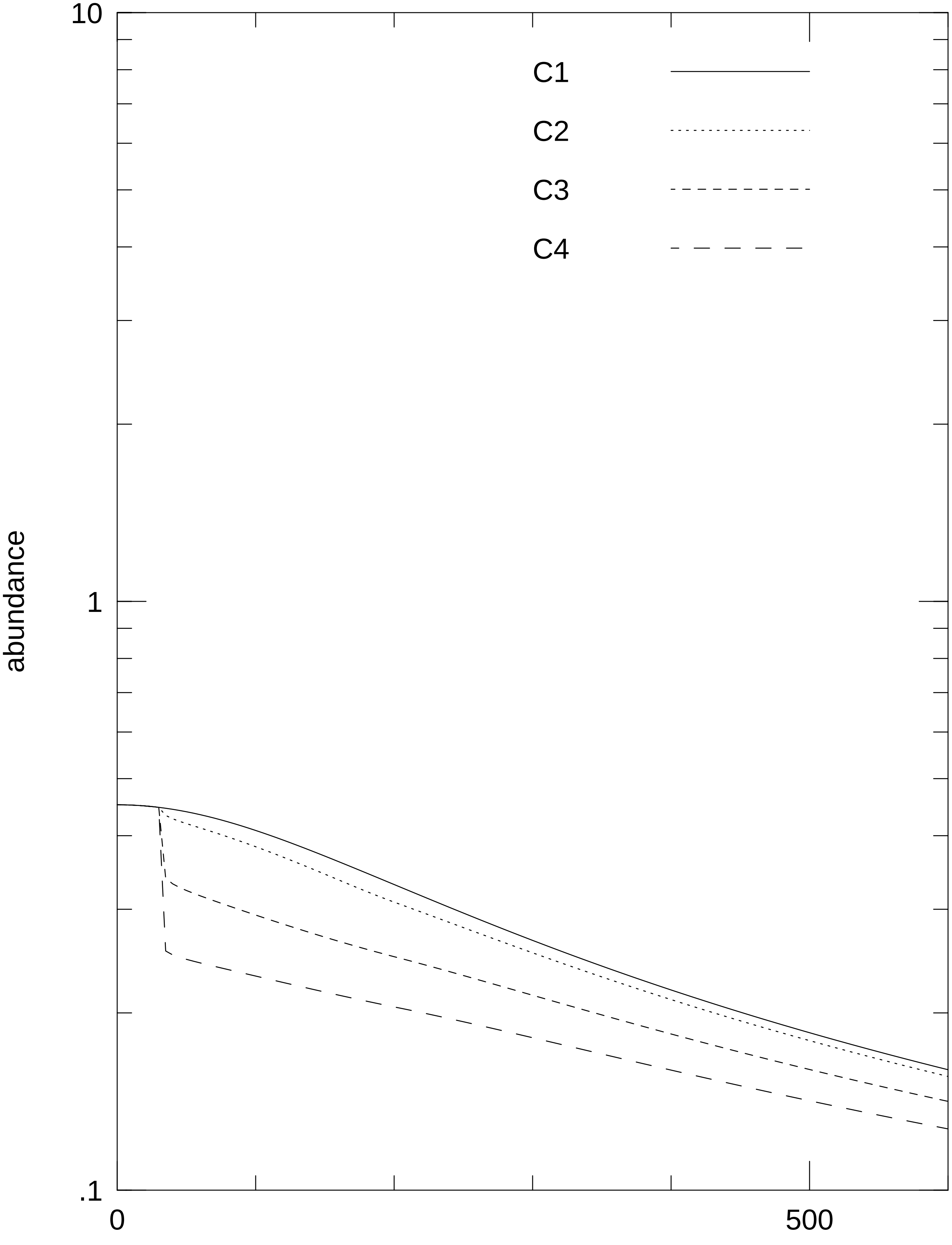

Fig.3

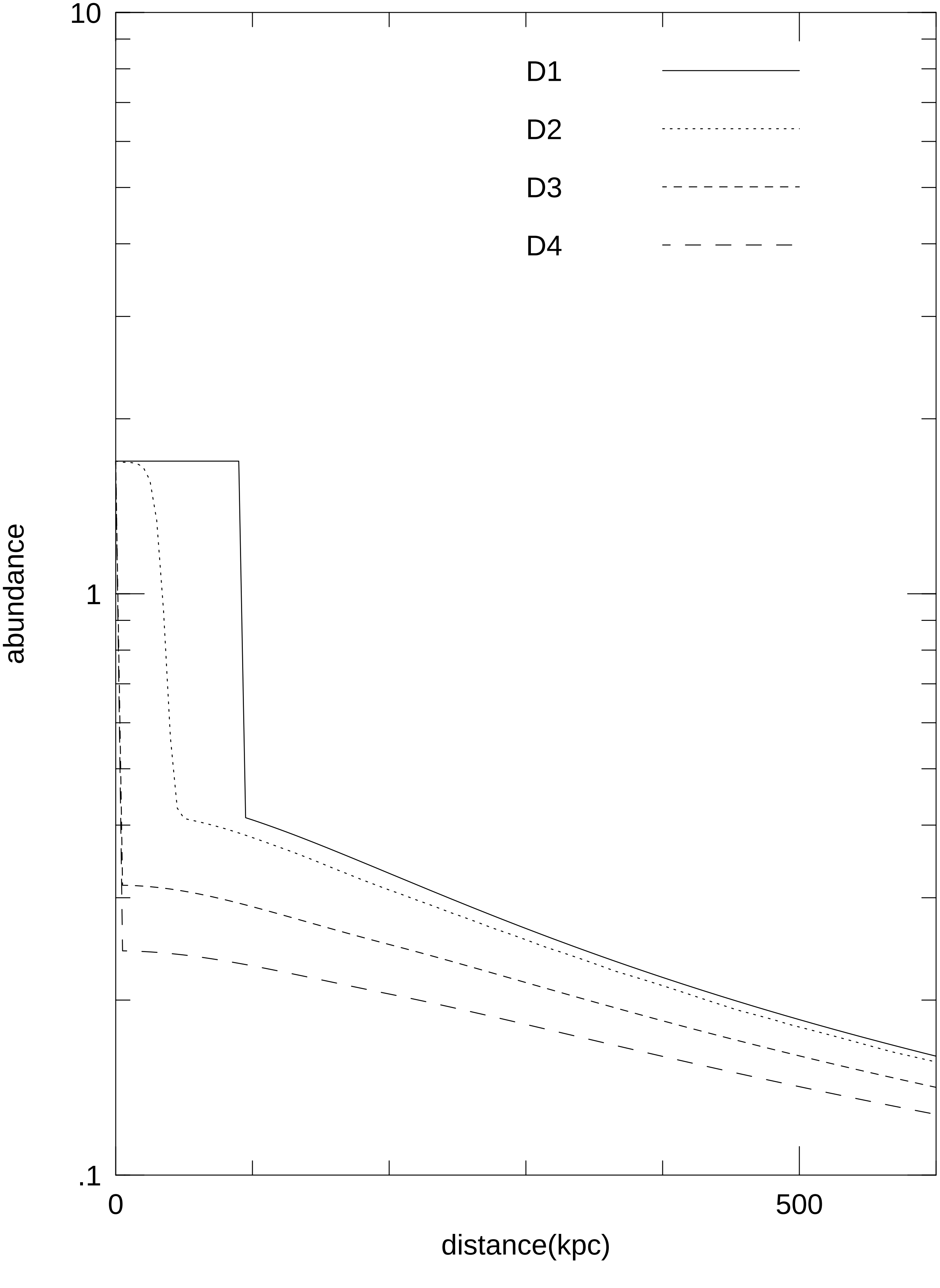

Fig.4

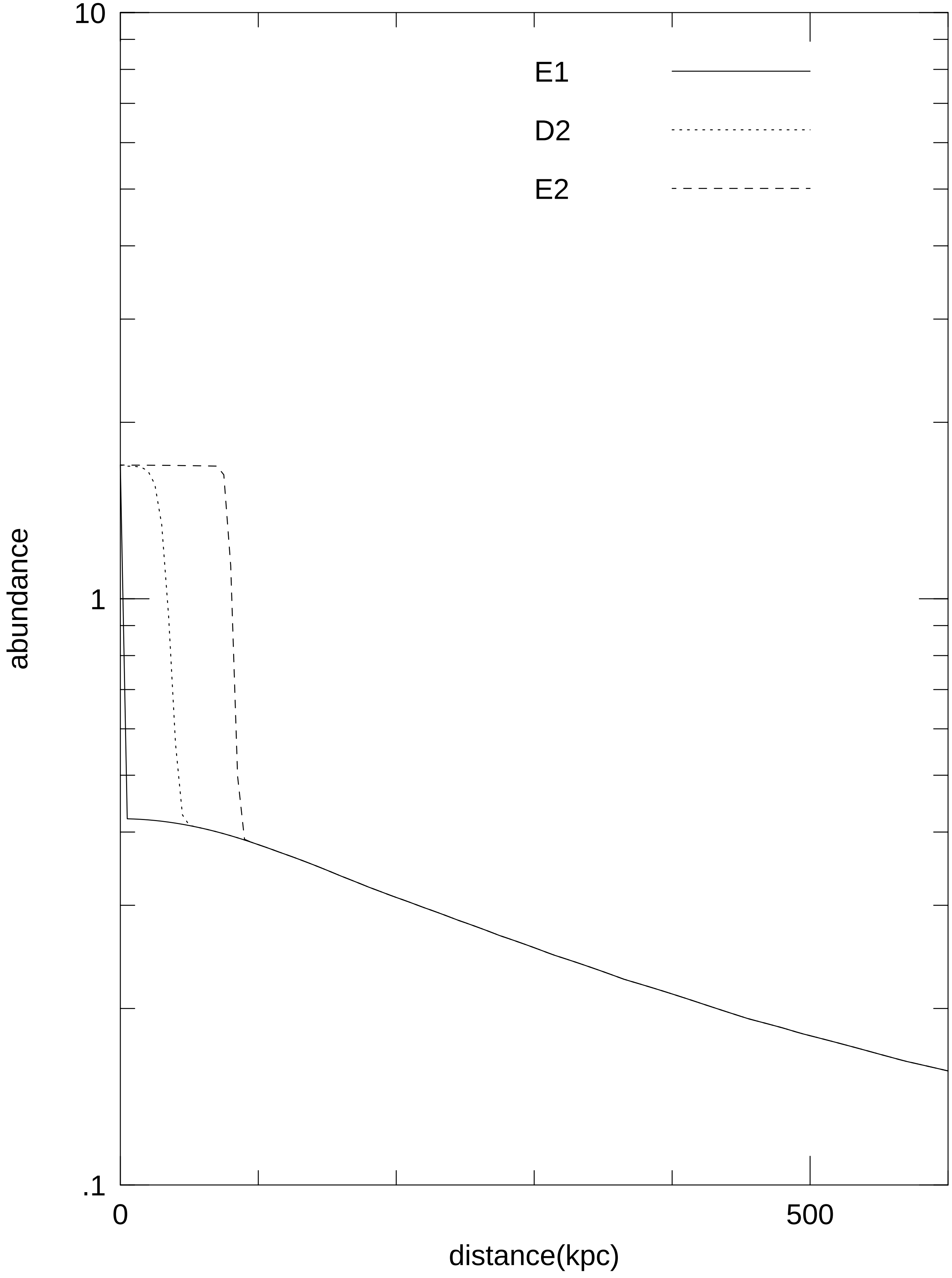

Fig.5(a)

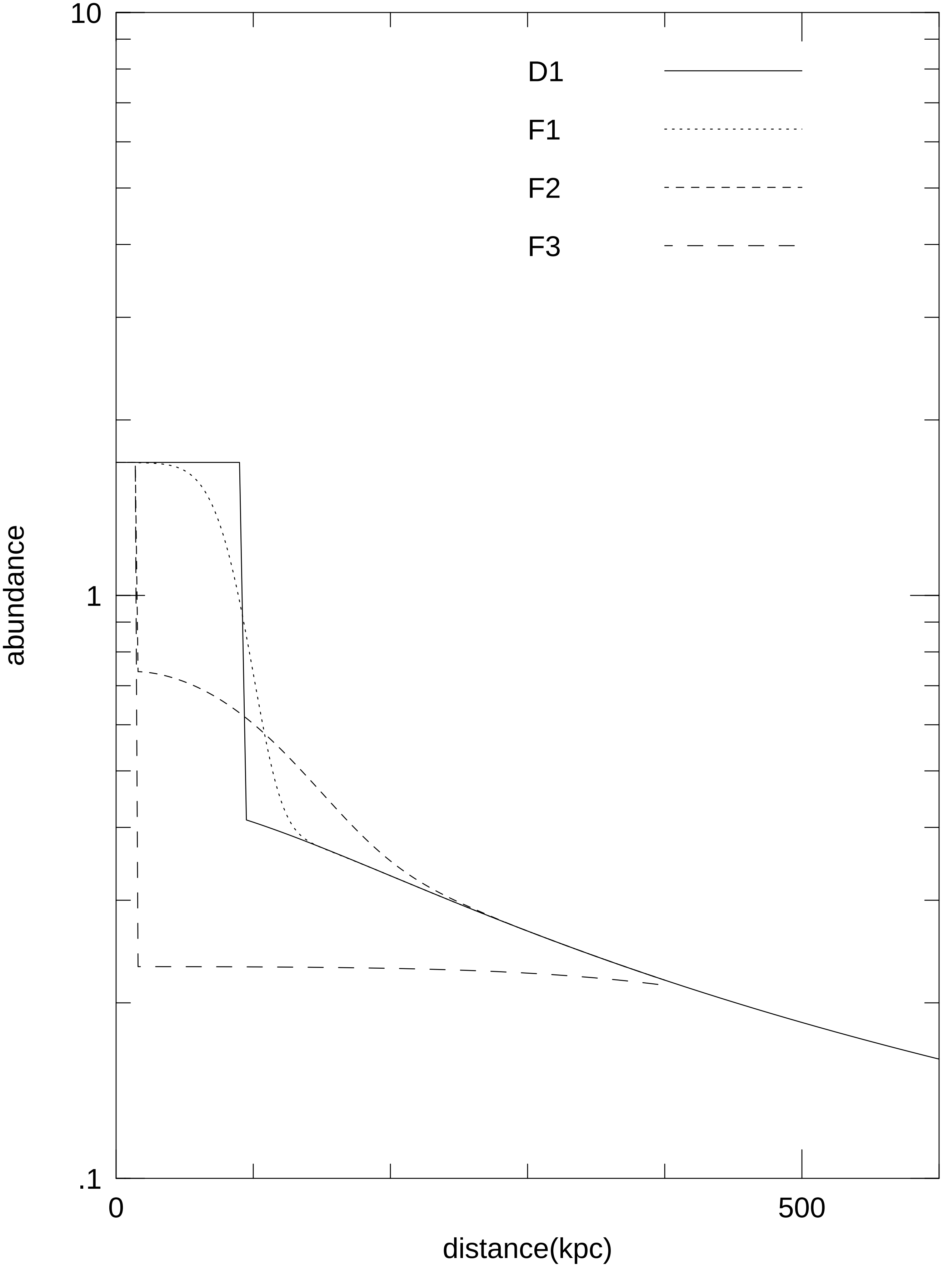

Fig.5(b)

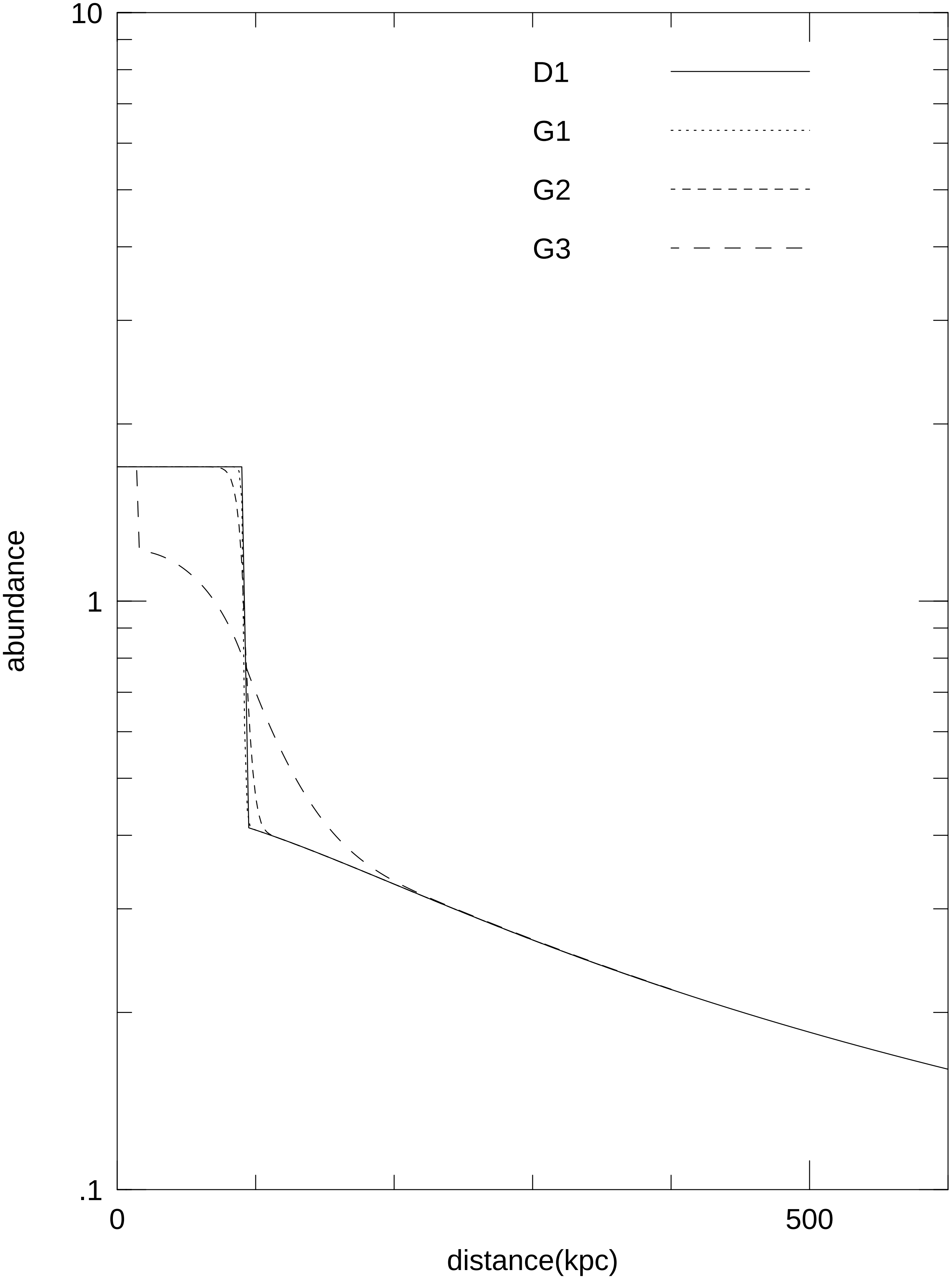

Fig.6(a)

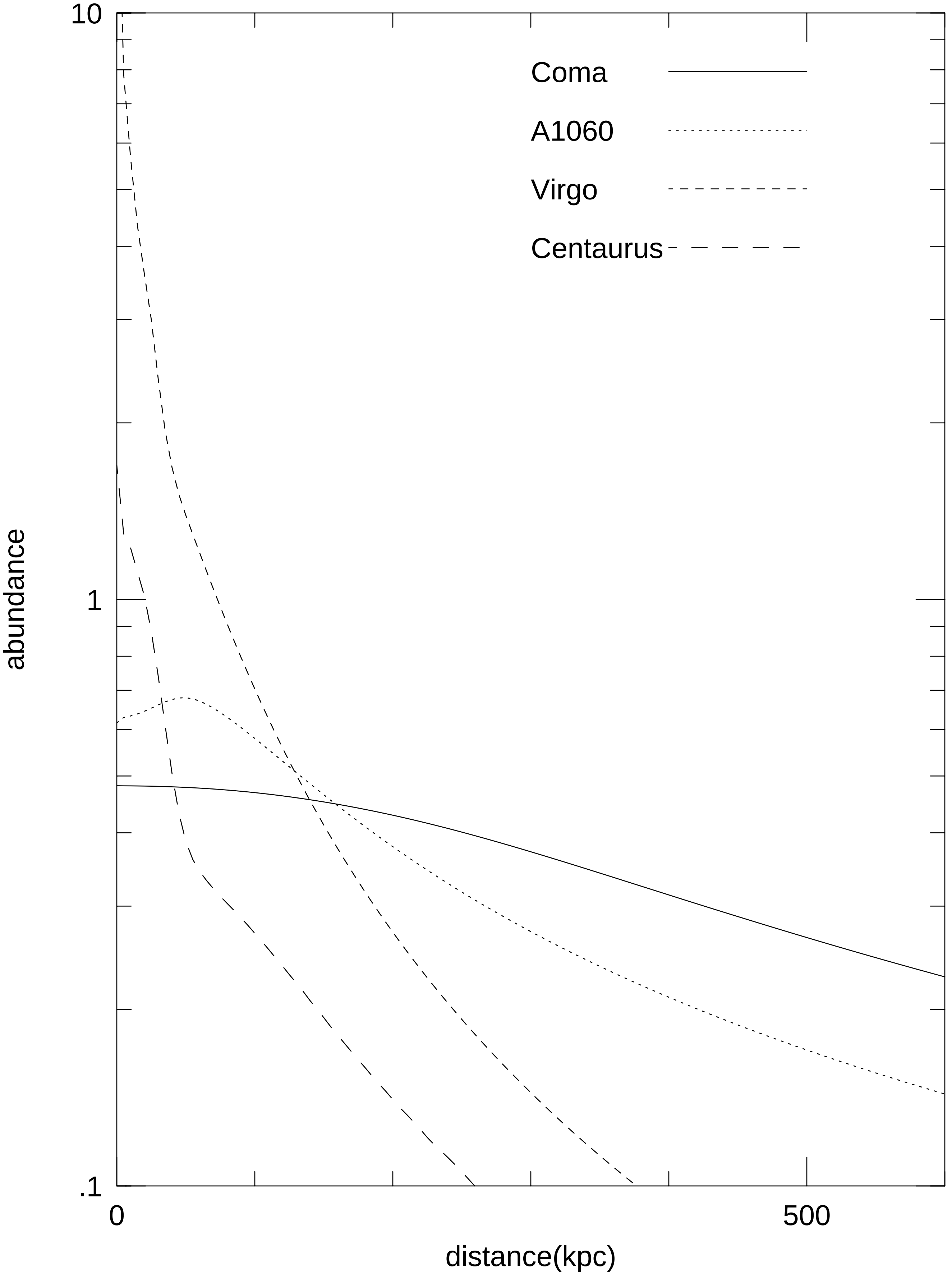

Fig.6(b)

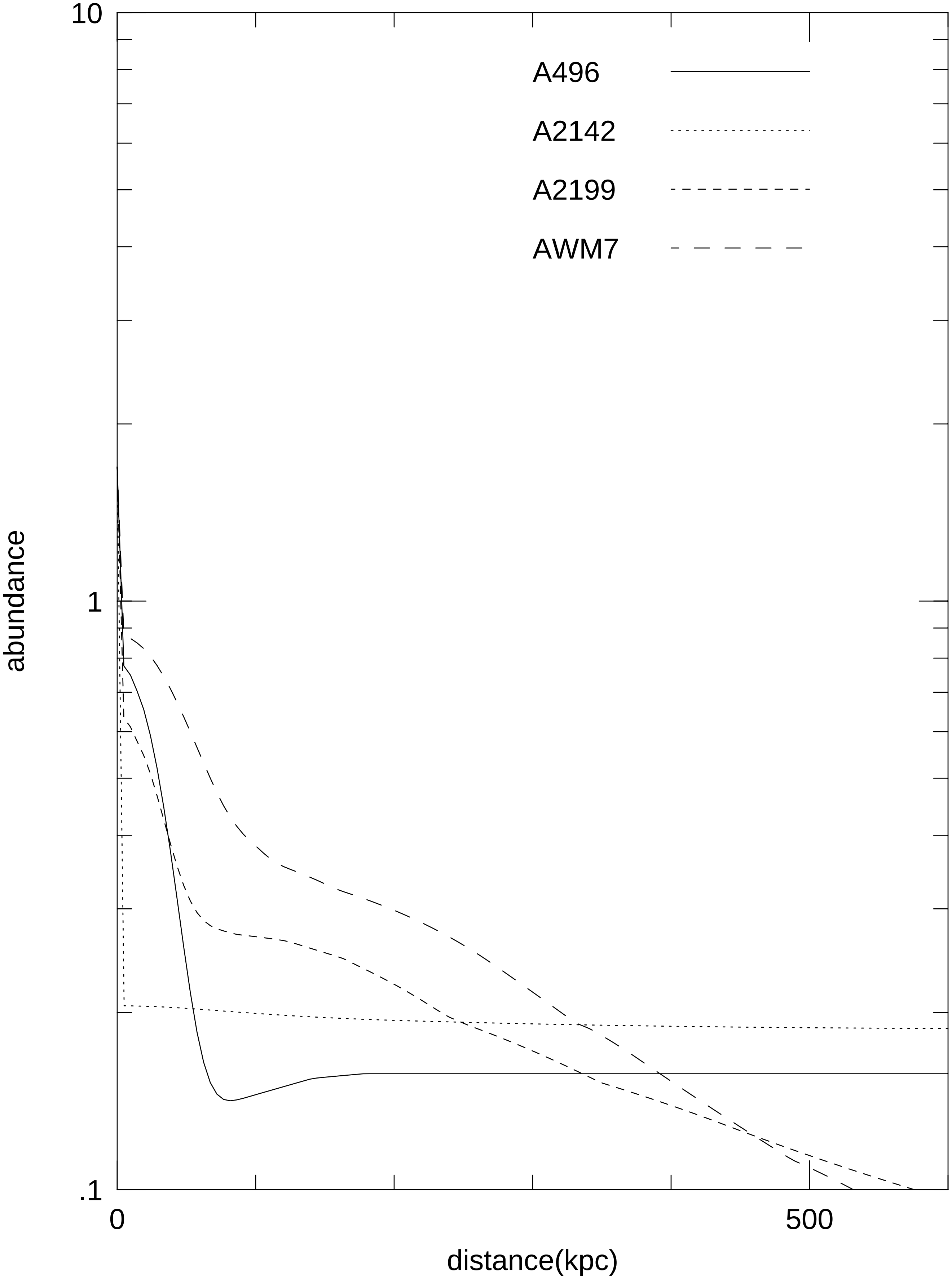

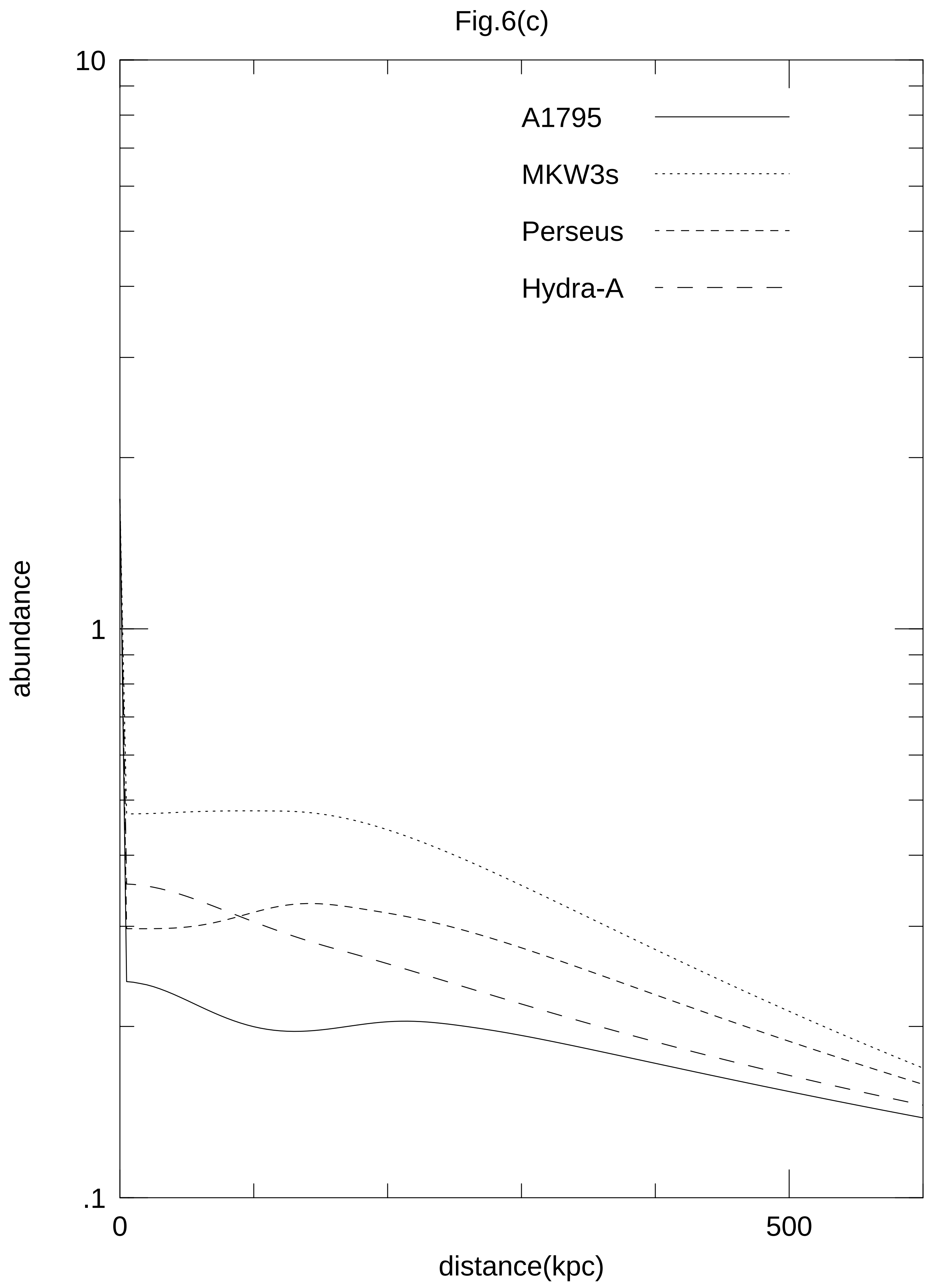

# Influence of Cooling Flow and Galactic Motion on the Iron Distribution in Clusters of Galaxies

Yutaka Fujita* and Hideo Kodama†

*Graduate School of Human and Environmental Studies
Kyoto University, Kyoto 606-01, Japan

*†Uji Research Center, Yukawa Institute for Theoretical Physics
Kyoto University, Uji 611, Japan

**Abstract**

Iron abundance distribution is now known for 12 clusters of galaxies. For some clusters (e.g. Centaurus) the observed abundance increases toward the cluster center, while for the others (e.g. Coma and Hydra-A) no significant inhomogeneity was observed. In order to understand this difference, we investigate the influence of cooling flow and turbulence produced by galactic motion on the iron abundance distribution by simple spherical models. We show that the cooling flow has a significant effect to flatten the iron abundance distribution if the flow velocity is sufficiently large. Further, by applying our analysis to the above clusters we show that we can give a systematic account for the observed variety of the iron abundance distribution qualitatively.

# 1 INTRODUCTION

It is now well established by X-ray observations that intracluster gas is contaminated by heavy elements up to the level of 1/4 to 1/2 of the solar abundance. However, until recently we have had little information on the spatial distribution of heavy elements inside each cluster of galaxies due to the poor spatial resolution of detectors. The exceptional cases were the Virgo, the Coma and the Perseus clusters. For the Virgo it was found that the iron abundance increases toward its center (Koyama, Takano, & Tawara 1991). In contrast for the Coma no significant inhomogeneity of the iron abundance was observed (Hughes, Gorenstein, & Fabricant 1988; Hughes et al. 1993). For the Perseus inhomogeneity of the iron abundance is a matter of argument (Ponman et al. 1990; Kowalski et al. 1993; Ohashi et al. 1994).

Some explanations for this difference in the iron abundance distribution have been presented so far. For the Virgo cluster Okazaki and others showed that the central concentration of iron is explained if one assumes that iron was ejected only from early type galaxies and that the iron distribution has not changed after the initial star burst (Okazaki et al. 1993). On the other hand for the Coma cluster Fabian et al. (1994) argued that its double-center structure suggests some violent merging phenomenon in its past, which smeared out the inhomogeneity of the iron distribution produced at its early phase.

The recent observation by ASCA (Ikebe et al. 1994 ; Fukazawa 1994 ; Fukazawa et al. 1994 ; Ohashi 1994) and the combined analysis of the Ginga LAC and Einstein SSS (White et al. 1994) have changed the situation of this problem significantly. We now have data on the spatial distribution of iron for 9 more clusters; Centaurus, AWM7, A496, A2142, A2199, A1060, A1795, MKW3s, and Hydra-A. The first five of them as well as Virgo have centrally concentrated distributions of iron, while for the rest no significant inhomogeneity was observed. (However White et al. show that the possibility of a weak iron concentration in A1795 cannot be completely discarded.) Though A1060 has the double-centered structure like Coma, the other clusters of the latter group has no such structure. So the above explanation for Coma does not apply to the latter group. Instead they have strong cooling flows. This suggests some relation between cooling flow and the iron abundance distribution.

Motivated by these observations, in the present paper, we investigate the influence of cooling flow on the iron abundance distribution by simple spherical models. We also examine the influence of the turbulent mixing provoked by the motion of galaxies by assuming that it can be effectively treated as diffusion.

The paper is organized as follows. First in the next section we describe the fundamental assumptions in detail and write down the basic equation to determine the time evolution of the iron abundance distribution. Then in §3 we solve this equation numerically for some typical models and quantitatively estimate the effect of cooling flow and motion of galaxies on the iron abundance distribution. On the basis of this general analysis, in §4, we construct specific models for each of the above 12 clusters taking into account the physical properties of the individual clusters, and by comparing the model predictions with observational data we discuss the possibility of accounting for the observed variety of the iron abundance distribution systematically. Section 5 is devoted to conclusion. Throughout this paper $H_0 = 50\mathrm{km}^{-1}\mathrm{s}^{-1}\mathrm{Mpc}^{-1}$ is assumed.

# 2 Assumptions and Basic Equations

We assume that the structure of a cluster of galaxies is described by spherically symmetric models. As a cooling flow affects only the central region of two or three core radii, we think that this approximation is good. Since the atomic diffusion of iron in the intracluster gas is negligible in the central region, under this assumption of spherical symmetry the time evolution of the iron abundance $Z(r,t)$ is described by the differential equation

$$\frac{\partial Z}{\partial t} = v\frac{\partial Z}{\partial r} + \frac{1}{r^2}\frac{\partial}{\partial r}\left(Dr^2\frac{\partial Z}{\partial r}\right) + S(r,t), \tag{1}$$

where $r$ is the distance from the cluster center, $v$ is the inward radial velocity, $D$ is an effective diffusion coefficient to be explained later, and $S(r,t)$ is the source term.

We consider two models as the source term in this paper. In the first model it is assumed that iron was first produced by star bursts in each galaxy at its formation and ejected into the intracluster space by galactic winds during some initial short period $0 \leq t \leq t_{\mathrm{GW}} (< 1\mathrm{Gyrs})$ approximately at a constant rate. Further it is assumed that iron was ejected only from early type galaxies in proportion to their masses (see Arnaud et al. 1992). Under these assumptions the source term is expressed as

$$S(r,t) = \frac{1}{t_{\mathrm{GW}}}\frac{f_{\mathrm{Fe}}}{1-f_{\mathrm{w}}}\frac{m_{\mathrm{p}}}{m_{\mathrm{Fe}}}f_{\mathrm{E}}(r)\frac{\rho_{\mathrm{gal}}(r)}{\rho_{\mathrm{gas}}(r)} \tag{2}$$

during the windy period and vanishes after that, where $\rho_{\rm gal}$ is the mass density of galaxies averaged over the dynamical time scale of the cluster, $\rho_{\rm gas}$ is the intracluster gas density, $m_{\rm p}$ is the proton mass, $m_{\rm Fe}$ is the atomic mass of iron, $f_{\rm E}(r)$ is the local fraction of early-type galaxies in a cluster, and $f_{\rm w}$ and $f_{\rm Fe}$ are the luminosity averaged mass fractions of gas and iron ejected from galaxies over the Hubble time, respectively.

In the cluster centers, however, it is very likely that additional iron was supplied from the big central cD galaxies and galaxies passing nearby at their formation. In the second model this effect is taken into account by modifying the above source model around the center in the following way.

First for simplicity we assume that the central cD galaxy affected the iron distribution only during the windy period, although it is at present controversial whether cD galaxies were formed within a short time after the formation of clusters and whether they affected other galaxies only during an early epoch (Malthmuth & Richstone 1984; Merritt 1984, 1985; Tonry 1987). Then, from the mass conservation, the metal rich gas ejected from the central cD galaxy and galaxies disrupted by it expands after the windy period to the region of the radius $R_{\rm cD}$ determined by

$$\frac{f_{\rm w}}{1-f_{\rm w}} \int_0^{R_{\rm cD}} 4\pi r^2 \rho_{\rm cD}(r) dr = \int_0^{R_{\rm cD}} 4\pi r^2 \rho_{\rm gas}(r) dr, \qquad (3)$$

where $\rho_{\rm cD}$ is the present stellar distribution of the cD galaxy. The iron abundance in this region is given by

$$Z_{\rm cD} = \frac{f_{\rm Fe}}{f_{\rm w}} \frac{m_{\rm p}}{m_{\rm Fe}}. \qquad (4)$$

Hence in the lowest approximation the source term for clusters with cD galaxies at center is given by

$$S(r,t) = \frac{1}{t_{GW}} Z_{\rm cD} \tag{5}$$

for $0 \le r \le R_{\rm cD}$, and by Eq.(2) outside. This is the second source model.

The spatial behavior of $S$ in the first model is determined by the mass distribution functions $\rho_{\rm gal}$, $f_{\rm E}$ and $\rho_{\rm gas}$. In the general analysis in §3 we assume that $\rho_{\rm gal}$ is given by the King model

$$\rho_{\rm gal}(r) = \rho_{\rm gal}(0)(1 + r^2/a^2)^{-3/2}, \tag{6}$$

and $\rho_{\rm gas}$ by the $\beta$-model

$$\rho_{\rm gas}(r) = \rho_{\rm gas}(0)(1 + r^2/a^2)^{-3\beta/2}, \tag{7}$$

and treat $\rho_{\rm gal}(0)$, $\rho_{\rm gas}(0)$ and the common core radius $a$ as free model parameters. As the fraction of elliptical galaxies two cases of $f_{\rm E} = 1$ and $f_{\rm E} = (e^{-r/a} + 1)/2$ are considered.

On the other hand, when we construct specific models of individual clusters in §4, we determine these parameters from observational data as far as possible. Further for some clusters different forms are adopted for $\rho_{\rm gas}$ from the observational ground. The details will be given in that section.

Throughout the present paper we assume that these density distributions do not change after the windy period. This assumption will be good for a region within two or three core radii from the center, provided that there occured no violent phenomena such as merging.

In the second model we must further specify $\rho_{\rm cD}$. Observations show that the surface brightness profiles of the halos of cD galaxies can be described by $r^{-2}$ (Oemler 1976). This is consistent with the result of the theoretical study by Duncan, Farouki and Shapiro (1983), who showed that mergers rapidly establish the $r^{-2}$ surface brightness profile outside the core of the cD galaxy. So it is natural to assume that the present star distribution of the cD galaxy, $\rho_{\rm cD}(r)$, is given by

$$\rho_{\rm cD}(r) = \rho_{\rm cD}(0)(1 + r^2/a_{\rm cD}^2)^{-3/2}. \tag{8}$$

The inhomogeneity in the iron distribution produced by the source term is modified by the cooling flow, which is represented by the first term on the right-hand side of equation (1). As is clear from this equation, they affect $Z$ only through the velocity of the cooling flow $v$.

In the cooling flow model $v(r)$ is related to the inward mass flow rate through the sphere of radius $r$, $\dot{M}(r)$, by

$$\dot{M}(r) = 4\pi r^2 \rho_{\rm gas}(r) v(r). \tag{9}$$

Using a multi-phase model, Thomas, Fabian, & Nulsen (1987) determined $\dot{M}(r)$ from observations and found that $\dot{M}(r)$ increases approximately in proportion to $r$ until around the cooling radius $r_{\rm c}$. Here the cooling radius is defined by the condition that the cooling time of the gas is equal to the Hubble time at $r = r_{\rm c}$. They also found that $\dot{M}$ vanishes at some finite radius from the center for some clusters. Though this feature is not definite enough due to the poor spatial resolution of observations, theoretical arguments suggest that it is very likely to be real (Fabian & Nulsen 1977; Cowie, Fabian, & Nulsen 1980, Loewenstein & Fabian 1990). Hence, in the present paper, we assume that $\dot{M}$ is expressed in the form

$$\dot{M}(r) = \begin{cases} 0 & \text{for } 0 \leq r \leq r_{\rm st} \ , \\ \dot{M}_{\rm c}(r - r_{\rm st})/(r_{\rm c} - r_{\rm st}) & \text{for } r_{\rm st} < r < r_{\rm c} \ , \\ \dot{M}_{\rm c} & \text{for } r_{\rm c} \leq r \ , \end{cases} \tag{10}$$

where $\dot{M}_{\rm c} = \dot{M}(r_{\rm c})$ and $r_{\rm st}$ are the parameter. We call $r_{\rm st}$ the stagnation radius.

Since we assume that $\rho_{\rm gas}(r)$ is time-dependent, the decrease of $\dot{M}(r)$ toward the center means that a substantial amount of hot gas is steadily lost from the cooling flow, which is balanced with a gas inflow from outer region. The most probable mechanism for this is the thermal instability in the flows (Nulsen 1986; Malagoli, Rosner, & Bongo 1987; White & Sarazin 1987; Balbus 1988; Balbus & Soker 1989). Since the thermal instability does not transfer heavy elements between hot regions and cool regions, we assume that the hot gas is removed from the cooling flow retaining its abundance.

We further assume that the cooling flow is time-independent and present after winds had stopped ($t > t_{\rm GW}$). Since the time scale of the winds is $\leq 10^9$yr (David, Forman, & Jones 1990, 1991a) and that of the cooling flow is $\sim 0.2 - 2 \times 10^{10}$yr (Edge, Stewart & Fabian 1992), this assumption seems to be reasonable.

In contrast to the bulk inflow motion, turbulence of hot gas produced by motion of galaxies and the associated diffusion or mixing of heavy elements are very intricate phenomena, and little theoretical investigation has been done on them. So we replace them by the following simple diffusion model in this paper. Each galaxy is assumed to disturb a spherical region of radius $R_{\rm eff}$, which we call the effective radius of the galaxy, and homogenizes the iron abundance inside it.

For any given point the abundance information diffuse over distance of order $R_{\rm eff}$, every time some galaxy passes by it within the distance $R_{\rm eff}$ from it. Hence, assuming that the motions of galaxies are random and isotropic locally, the effective diffusion coefficient $D$ is given by

$$D \simeq R_{\rm eff}^2 \times n_{\rm gal}\sigma\pi R_{\rm eff}^2/6 = \pi\sigma n_{\rm gal}(r) R_{\rm eff}^4/6, \tag{11}$$

where $\sigma$ and $n_{\rm gal}(r)$ are the local virial velocity and the local number density of galaxies. We assume that the $r$ dependence of $n_{\rm gal}(r)$ is given by the King model Eq.(6) with $\rho_{\rm gal}(0)$ replaced by $n_{\rm gal}(0)$. If the galactic winds are not so strong, $R_{\rm eff}$ is expected to be at most of the order of the tidal radius of the cluster gravitational field,

$$R_{\rm eff} \leq 60(a/500{\rm kpc}){\rm kpc} \tag{12}$$

(Malthmuth & Richstone 1984; Merritt 1984).

# 3 GENERAL NUMERICAL ANALYSIS

The generic model explained in the previous section has 15 parameters apart from the freedom in the choice of $f_{\rm E}$; $t_{\rm GW}$, $f_{\rm Fe}$, $f_{\rm W}$, $\rho_{\rm gal}(0)$, $\rho_{\rm gas}(0)$, $a$, $\beta$, $\rho_{\rm cD}(0)$, $a_{\rm cD}$, $\dot{M}_{\rm c}$, $r_{\rm st}$, $r_{\rm c}$, $\sigma$, $n_{\rm gal}(0)$ and $R_{\rm eff}$. In order to determine the present value of $Z$ by integrating the evolution equation, we must also specify the time from the beginning of the windy period, $t_0$. However, these parameters do not affect the final value of $Z$ independently. In fact, if we introduce the dimensionless quantities

$$\tau = t/t_0, \tag{13}$$

$$x = r/a, \tag{14}$$

$$\tilde{v} = vt_0/a, \quad (15)$$

$$\tilde{D} = Dt_0/a^2, \quad (16)$$

$$\tilde{S} = St_0, \quad (17)$$

the evolution equation (1) is written only in terms of them as

$$\frac{\partial Z}{\partial \tau} = \tilde{v}\frac{\partial Z}{\partial x} + \frac{1}{x^2}\frac{\partial}{\partial x}\left(\tilde{D}x^2\frac{\partial Z}{\partial x}\right) + \tilde{S}. \quad (18)$$

Here the $x$-dependence of the dimensionless quantities $\tilde{v}$, $\tilde{D}$ and $\tilde{S}$ are given by

$$\tilde{v}(x) = \begin{cases} \tilde{v}_a x^{-2}(1+x^2)^{3\beta/2}\frac{x-r_{\rm st}/a}{r_{\rm c}/a-r_{\rm st}/a} & (r_{\rm st}/a < x < r_{\rm c}/a) \\ \tilde{v}_a x^{-2}(1+x^2)^{3\beta/2} & (r_{\rm c}/a < x) \end{cases} \quad (19)$$

$$\tilde{D}(x) = \tilde{D}_0(1+x^2)^{-3/2}, \quad (20)$$

$$\tilde{S}(x,\tau) = \frac{t_0}{t_{\rm GW}} \times \begin{cases} Z_{\rm cD} & (x < R_{\rm cD}/a) \\ Z_0 f_{\rm E}(x)(1+x^2)^{3(\beta-1)/2} & (R_{\rm cD}/a < x) \end{cases}, \quad (21)$$

where

$$\tilde{v}_a = 2.8\times 10^{-2}\frac{t_0}{1.33\times 10^{10}{\rm yr}}\frac{\dot{M}_{\rm c}}{100M_\odot/{\rm yr}} \times \left(\frac{\rho_{\rm gas}(0)}{10^{-2}m_{\rm p}/{\rm cm}^3}\right)^{-1}\left(\frac{a}{250{\rm kpc}}\right)^{-3}, \quad (22)$$

$$\tilde{D}_0 = 7.3\times 10^{-4}\frac{t_0}{1.33\times 10^{10}{\rm yr}}\frac{\sigma}{10^3{\rm km/s}}\frac{a^3 n_{\rm gal}(0)}{10} \times \left(\frac{a}{250{\rm kpc}}\right)^{-5}\left(\frac{R_{\rm eff}}{10{\rm kpc}}\right)^4, \quad (23)$$

$$Z_0 = Z_{\rm cD}\frac{f_{\rm W}}{1-f_{\rm W}}\frac{\rho_{\rm gal}(0)}{\rho_{\rm gas}(0)}. \quad (24)$$

From these equations we see that only 11 of the 16 parameters are independent in effect.

In this section we reduce the number of these free parameters by giving typical fixed values to some of them, because the main purpose of this section is to look into the characteristic features of the influence of the cooling flow and the galactic motion on the iron distribution evolution and get rough estimates of it.

First we take $a = r_c = 200$kpc, $t_0 = 10^{10}$yr, $t_{GW} = 10^8$yr, and $\rho_{gas} = 5 \times 10^{-3} m_p/\mathrm{cm}^3$. For this choice the dimensionless parameter $\tilde{v}_a$ is controlled by $\dot{M}_c$.

Next in order to fix the source term we take $f_w = 0.18$ and $f_{Fe} = 6.8 \times 10^{-4}$, which correspond to the IMF exponent $x = 1.5$ (David et al. 1991b). Although David et al. (1991b) showed from the comparison with observations that $x = 1.0$ is more preferable than $x = 1.5$, Okazaki et al. (1993) pointed out that the ejected iron masses in Davids' models are rather overestimated because the iron which is used in star formation in stars is neglected in their models. Therefore we adopt the exponent $x = 1.5$. For this choice $Z_{cD} = 6.8 \times 10^{-5}$, which is 1.7 times the solar iron abundance $4 \times 10^{-5}$. To determine $\rho_{gal}(0)$ we use the empirical relation between the number of the galaxies within 500kpc from the cluster center, $N_{500}$, and the total luminosity of them. For $M/L = 8M_\odot/L_\odot$ this relation is expressed in terms of the total mass of galaxies in the same region, $M_{500}$, and $N_{500}$ as

$$M_{500} = 1.44 \times 10^{11} N_{500}^{1.42} \mathrm{M}_\odot. \quad (25)$$

(Edge & Stewart 1991b). With this relation and Eq. (6) $\rho_{gal}(0)$ is determined. We tentatively take $N_{500} = 21$ (Bahcall 1981) or $\rho_{gal}(0)/m_p = 6.1 \times 10^{-3}\mathrm{cm}^{-3}$, which fixes $Z_0$ to be $1.8 \times 10^{-5}$. Finally we take $n_{cD}(0)(= \rho_{cD}(0)/m_p) = 180\mathrm{cm}^{-3}$ and $a_{cD} = 2$kpc (Malumuth & Kirshner 1985) except for model E explained below for which $n_{cD}(0)$ is varied. This corresponds to the velocity dispersion of 330 km/s from equation $\sigma_{\rm cD}^2 = 4\pi G\rho_{\rm cD}(0)a_{\rm cD}^2[3/2 - 2\ln 2]$ (Tonry 1985), where $\sigma_{\rm cD}$ is the velocity dispersion of stars in the cD galaxy. These parameters fixes the value of $R_{\rm cD}$. Thus there is no free parameter in the source model(except for model E).

Finally in the diffusion coefficient we take $\sigma_r = 1000$km/s where $\sigma_r$ is the line-of-sight velocity dispersion or $\sigma_r = \sigma/\sqrt{3}$, and normalize $n_{\rm gal}(0)$ so that $N_{500} = 21$ as above. This makes $R_{\rm eff}$ the control parameter of $\tilde{D}_0$.

For this parameter choice we are left with 5 free parameters, $\dot{M}_{\rm c}$, $\beta$, $r_{\rm st}$, $R_{\rm eff}$, and $n_{\rm cD}(0)$. In order to see the effect of these parameters on the iron abundance distribution, we solve the evolution equation (18) for a series of models whose details are given in Table 1.

Models labeled by A, B, C, D and E are ones to see the effect of cooling flows, for which $D = 0$. In the series A, B, C the first source model is used as $S$, while in the others the second source model is adopted. $f_{\rm E}$ is put $(e^{-r/a} + 1)/2$ in B models to see its effect but is put unity for the others. Models A and C are built to see the effect of $\beta$ and $r_{\rm st}$. Since the fitted values of $\beta$ are contained in the range $0.5 \leq \beta \leq 0.7$ for most clusters, we take 0.5 and 0.7 as typical values. Observationally $\rho_{\rm cD}(r)$ $(r >> a_{\rm cD})$ of various clusters determined by velocity dispersions and core radii are in the range of factor 2 or 3 of the mean value in the 1 sigma level(Table 5 of Malumuth & Kirshner 1985). Model E is to see the variation caused by this scattering. $a_{\rm cD}$ is fixed to be 2kpc.

On the other hand models F and G are taken to see the effect of galactic motions. Their model parameters are the same as those of model D except that the diffusion term is effective. The difference between models F and G are the length $t_{\rm d}$ of the period when the diffusive mixing is effective: $t_{\rm d} = 10^{10}$yr $= t_0$ in model F and $t_{\rm d} = 10^8$yr $= t_{\rm GW}$ in model G. In both models we switch off the cooling flow to see the genuine effect of galactic motion. Further we set $D(r) = 0$ for $r < 17$kpc for the following reason. Galaxies going through this region will be disrupted by tidal force of the cD galaxy. As a result these galaxies do not provoke mixing, but instead contribute to the source term, which is taken into account in the second source model. The tidal radius of a galaxy with mass $m$ at distance $r$ from the cluster center is expressed in terms of the cluster mass inside that radius, $M(r)$ as

$$R_{\rm t} = (m/3M(r))^{1/3} r \tag{26}$$

(Binney & Tremaine 1987). Since $M(r)$ is dominated by the cD galaxy mass in the present case, for the typical values $m = 5 \times 10^{11} M_\odot$ and $R_{\rm t} = 10$kpc, this equation yields $r \approx 17$kpc.

The value of $Z$ at $t = t_0$ in units of the solar abundance $4 \times 10^{-5}$ for these models obtained by numerical integration of Eq.(1) are shown in Figure 1 to 5. Further in Table 1 the two characteristic dimensionless velocities of the models, $\tilde{u}_a$ and $\tilde{u}_{R_{\rm cD}}$, and the abundance concentration parameter $\tilde{Z}$ defined as follows are shown:

$$\tilde{u}(r) = v(r) t_0 / r, \tag{27}$$

$$\tilde{u}_{RcD} = \tilde{u}(R_{\rm cD}), \tag{28}$$

$$\tilde{u}_a = \tilde{u}(a), \tag{29}$$

$$\tilde{Z} = Z(0.1a, t_0)/Z(1.0a, t_0). \tag{30}$$

Figure 1*a*, 1*b* and Table 1 show that the cooling flow flattens the initial iron inhomogeneity produced by the source term quite efficiently if the flow velocity is sufficiently large. The reason of this flattening is quite simple: it is an inward drift and compression of the distribution pattern by the cooling flow. To be precise, the value of $Z$ is conserved in the Lagrange coordinate of the flow when the diffusion is negligible after the windy period. Since the iron abundance distribution just after the windy period becomes flatter in the outer region as $r$ increases, the final distribution becomes flatter when the characteristic flow velocity $\tilde{u}_a$ increases, as is seen in Table 1.

The cooling flow does not affect the value of $Z$ at the center where $\dot{M}$ vanishes. This produces the sharp peaks at the center in Figures 1a, 1b and 2a. For the same reason we get central peaks of width $r_{\rm st}$ in model C as is clearly seen in Figure 2b. The inverse correlation between $\tilde{u}_a$ and $\tilde{Z}$ for model C in Table 1 is due to these central peaks. If these peaks are observed in clusters with flat iron abundance distributions and strong cooling flows, they will strongly suggest that the flat distributions have been produced by cooling flows. However, as these peaks have narrow widths, they may not be observed by detectors with low spatial resolutions even if they exist.

As is seen from comparison of Figures 1a, 1b and 2a, the flattening effect of cooling flow becomes more prominent as the source term is more centrally concentrated for the source model with no cD contribution. In spite of this source term dependence Table 1 shows that the initial inhomogeneity produced by the source term is practically erased if $\tilde{u}_a$ is larger than unity.

The abundance distribution just after the windy period in the source model with cD contribution is different from that in the source model without cD contribution due to the existence of a broad, high central peak. This peak is made narrower by cooling flow as shown in Figure 3. As is seen in Figure 4, the final width of the peak is very sensitive to the value of $n_{\rm cD}(0)$. This is because the initial width of the peak, $R_{\rm cD}$, increases with $n_{\rm cD}(0)$ and $\tilde{u}(r)$ decreases with $r$. Note that in spite of this sensitivity to physical parameters the phenomenological quantity $\tilde{u}_{R_{\rm cD}}$ yields a good index to decide the fate of the peak. Roughly speaking, the peak disappears if the value of $\tilde{u}_{R_{\rm cD}}$ is around unity or greater.

The final distributions in these models D and E resemble those in model C with $r_{\rm st} \neq 0$ if the cooling flow is moderate. However, since the peak value is larger in the former than in the latter, they can be distinguished by observations with good spatial resolutions.

Finally Figure 5 and Table 1 show that the efficiency of the diffusive mixing due to galaxy motion is sensitive to $R_{\rm eff}$ and the duration of the diffusion. In particular from Figure 5*a* we see that the central peak due to the cD contribution is smeared out except in the region $r < R_{\rm t}$ if $R_{\rm eff}$ is larger than 10kpc during most of the cluster life. The cooling flow can easily erase this residual peak within $R_{\rm t}$ if it reaches the cluster center. As is discussed in detail in the next section, however, it is likely that such strong diffusion does not occur in reality. Figure 5*b* and Table 1 also show that the central peak by the cD contribution survives after the windy period even if $R_{\rm eff}$ becomes as large as 30kpc due to galactic winds.

# 4 SPECIFIC MODELS OF THE OBSERVED CLUSTERS

As stated in the introduction, among the 12 clusters for which the iron abundance distribution has been observed, Virgo (Koyama et al. 1991), Centaurus (Fukazawa et al. 1994), AWM7 (Ohashi 1994; Fukazawa 1994), A496, A2142 and A2199 (White et al. 1994) have abundance gradients, but in Coma (Hughes et al. 1988, 1993), A1060, MKW3s (Ohashi 1994; Fukazawa 1994), Hydra-A (Ikebe et al. 1994), and A1795 (White et al. 1994), the gradients are small. Ponman et al (1990) and Kowalski et al (1993) reported the presence of the abundance gradient in the Perseus cluster. However, the recent ASCA observation gave a contrary result(Ohashi 1994). The physical parameters of these clusters determined by observations are summarized in Table 2 where $n_\beta$ and $n_{\rm d}$ are explained later.

First from these data we construct specific models for each cluster within the framework given in §2.

As the galaxy distribution function $\rho_{\rm gal}$ we use the King model (6) for all clusters except for Virgo cluster and determine $\rho_{\rm gal}(0)$ for each cluster with the help of Eq. (25). We assume the core radii of Eq. (6) are the same as those of Eq. (7) except for AWM7 and MKW3s. The cooling radii of AWM7 and MKW3s are determined from Figure 1 of Canizares, Stewart & Fabian (1983). Since $N_{500}$ is not known for MKW3s and Hydra-A, we tentatively take $N_{500} = 21$ (Bahcall 1981) for these clusters. Finally for Virgo cluster, since the cooling flow should be regard as that of M87, we use the stellar mass density as $\rho_{\rm gal}(r)$.

For most clusters $\rho_{\rm gas}(r)$ is well fitted by the $\beta$ model (7) outside the cooling radius, but for clusters with strong cooling flows the best fit $\beta$ models yield smaller values in the central region than the observed values (Jones & Forman 1984). Since this deviation affects the cooling flow velocity, we modify the simple $\beta$ model as follows for $r < r_{\rm c}$:

$$\rho_{\rm gas}^{\rm c}(r) \quad = \quad (n_{\rm d}/n_\beta)^{1-r/r_{\rm c}} \rho_{\rm gas}(r)\,, \tag{31}$$

where $n_\beta$ is $\rho_{\rm gas}(0)/m_{\rm p}$ of the best fit $\beta$ model $\rho_{\rm gas}(r)$, and $n_{\rm d}$ is the corresponding observed value.

Further for AWM7 and MKW3s the gas density is not be fitted by the $\beta$ model. Instead the following simple power-law model yields a good approximation(Kriss, Cioffi, & Canizares 1983):

$$\rho_{\rm gas}(r) = m_{\rm p} n_{\rm b} (r/1{\rm kpc})^\alpha. \tag{32}$$

The best fit values of the parameters are $n_{\rm b} = 0.32{\rm cm}^{-3}$ and $\alpha = -1.04$ for AWM7, and $n_{\rm b} = 2.17{\rm cm}^{-3}$ and $\alpha = -1.35$ for MKW3s. In these two clusters, the mass deposition rate is given at the radius of the central bin used in the observations, which are 70kpc for AWM7 and 40kpc for MKW3s (Canizares, Stewart, & Fabian 1983). Taking account of these spatial resolutions of the observations, we modify Eq.(32) so that $\rho_{\rm gas}$ is constant within these central bins in order to avoid the divergence at the center.

For all the clusters we assume $t_0 = 10^{10}$yr and $t_{\rm GW} = 10^8$yr. We also take $f_{\rm w} = 0.18$ and $f_{\rm Fe} = 6.8 \times 10^{-4}$ as in the previous section.

For Coma and A1060 we apply model A in the previous section because, though there are two cD galaxies in these clusters, they are not at rest at the centers. With $f_{\rm w}$ and $f_{\rm Fe}$ given above, the density distribution models explained above completely fixes the source term for these clusters.

On the other hand for the other clusters we construct models by adding the diffusive mixing term by the galactic motion to model D. We use the fixed common values, $a_{\rm cD} = 2$kpc and $\rho_{\rm cD}(0)/m_{\rm p} = 180{\rm cm}^{-3}$, to calculate the contribution from cD galaxies to the source term for all the clusters as we have no data on them.

The study of Bahcall (1981) shows that there is no noticeable difference in the line-of-sight velocity dispersion of galaxies between the clusters with centrally concentrated iron distributions and the others: its mean value is 854km/s for the former clusters, Virgo, Centaurus, A2142, A2199 and AWM7, and 960km/s for the latter, Coma, A1060, A1795 and Perseus. If the observed central concentrations are assumed to be produced by the central cD galaxies, this and the result for model F in the previous section suggest that $R_{\rm eff}$ must be smaller than 10kpc around the center during most of the lives of the clusters, provided that the value of $R_{\rm eff}$ does not vary from cluster to cluster. This is consistent with the facts that the elliptical fraction is high in the central region of clusters (e.g. Whitmore & Gilmore 1993) and elliptical galaxies have less gas than spiral galaxies, because a completely stripped galaxy hardly disturbs intracluster gas as shown by Gaetz, Salpeter and Shaviv (1987) (see Fig. 2$a$ in their paper). Therefore we take as $R_{\rm eff}$ the value determined by the right hand of Eq.(12) for $0 < t < t_{\rm GW}$ and 5kpc for $t_{\rm GW} < t$. This with the observed values of $\sigma$ and $n_{\rm gal}$ determine the effective diffusion coefficient. As in the previous section $n_{\rm gal}(0)$ is determined by $N_{500}$ (Bahcall 1981) and we assume $D(r) = 0$ for $r < 17$kpc.

The present profiles of the iron abundance distributions calculated for these specific models are shown in Figure 6(a) - 6(c). The corresponding characteristic velocities $\tilde{u}_a$ and $\tilde{u}_{R_{\rm cD}}$ and the final density contrast index $\tilde{Z}$ are listed in Table 3. Here $\tilde{Z}$ is defined by Eq.(30) except for Virgo while the value defined by

$$\tilde{Z} = Z(2', t_0)/Z(10', t_0) \tag{33}$$

is given for Virgo, assuming the distance to it is 15Mpc. The reason we take the values at $2'$ and $10'$ is that the iron abundance gradient is actually observed in this region by ASCA (Koyama private communication).

Figure 6(a) - 6(c) and Table 3 show that our model calculation gives iron abundance distributions which are consistent with the observations for the clusters observed by ASCA (Coma, 1060, Virgo, Centaurus, AWM7, MKW3s, Perseus and Hydra-A) and A1795. This indicates that the combination of the strength of the cooling flows and the contribution of the central cD galaxies explains the variety of the iron abundance distributions for these clusters. In particular it can be understood that the cooling flows of Virgo, Centaurus and AWM7 have been too weak to modify the initial iron distributions while they have been erased by strong cooling flow compression in A1795, MKW3s, Perseus and Hydra-A (see $\tilde{u}_{RcD}$ of Table 3).

On the other hand the situation for A496 and A2199 is subtle. Though our calculation gives central peaks for them, they are almost within $1'$. Hence when they are averaged over $6'$ the angular scale corresponding to the field of view of Einstein SSS, they may not give as large central enhancement as observed. Finally in the case of A2142 the central peak disappears in our model contrary to the observation. As we mentioned in §3, however, there is an uncertainty in $n_{\rm cD}(0)$ and the duration of the cooling flow, so the result is not conclusive.

One interesting point of our results is that the uniform iron distribution of Coma and A1060 are accounted for even if no merging has occured. Of course this does not imply that possible merging phenomena in the past are not relevant to the present iron abundance distribution. Instead it rather indicate that the contribution of the central cD galaxy is crucial in accounting for the present central concentration of the iron in clusters with central cD galaxies because if the contribution were neglected, we would obtain flat distributions as in Coma and A1060.

# 5 CONCLUSION

In the present paper we have investigated the influence of cooling flow and galactic motion on the iron abundance distribution in clusters of galaxies by simple spherical models. We have performed numerical calculations for some typical models and shown that the cooling flow is very effective in flattening the centrally concentration of iron abundance produced during the initial windy period if the flow velocity is sufficiently large. We have also shown that the supply of the metal rich gas by the central cD galaxies are essential in accounting for the existence of clusters with highly central concentrated distribution of iron abundance, and that the diffusive mixing by galactic motion should be small enough for those central peaks to survive. The latter is consistent with the fact that the central regions of clusters are usually dominated by gas-poor elliptical galaxies, but gives the constraint that the effective radius of galaxies should be smaller than 30kpc during the windy period.

Further on the basis of this general analysis we have constructed specific models for each of the observed clusters. We have shown that the observed variety in the iron abundance distribution of clusters with central cD galaxies at rest at center can be qualitatively explained in our scheme as a result of the production of a centrally concentrated iron distribution by the cD galaxy and its subsequent deformation by cooling flow, except for a few clusters for which the available data have too poor resolution or some important physical parameter are unknown. This result is remarkable taking into account the simplicity of our model, and we believe that it suggests the validity of our scenario.

One important feature of our model is the existence of a clear observable prediction. Since the flat distribution of the iron abundance is produced by the central compression of the iron rich region toward the center in our model, a sharp peak in the iron abundance distribution is left at the center if the cooling flow does not reach the center. If such a peak is detected by future observation with high spatial resolutions, it will give a strong evidence for the validity of our scenario.

Of course such a peak may not be left if the effective radii of galaxies become larger than 30kpc during the windy period. If we extend our argument to other heavy elements with this constraint in mind, we obtain another interesting prediction. According to the study of the chemical evolution of galaxies (e.g. David et al. 1991b) strong galactic winds are produced by Type II supernovae which eject oxygen rich gas, while iron enriched gas is ejected by Type I supernovae at low ejection rate after the strong wind period. Thus, if our scenario is correct, this leads to the prediction that the iron distribution has a central peak but the oxygen abundance is uniform in clusters of galaxies which have central cD galaxies at the centers and have weak cooling flows. The observational check of the prediction will give useful information on the structure of galaxies during the windy period as well as their chemical evolution.

We would like to thank K.Koyama, A.Habe and T.Tsuru for useful suggestions and providing us new data of ASCA. We also wish to thank K.Tomita, J.Yokoyama, Y.Yamada, R.Nishi, Y.Fujiwara and N.Makino for discussion.

# Figure Captions

Fig.1.- Iron abundance distributions in units of the solar abundance as a function of the distance from the cluster center. (a) model A1. (b) model A2.

Fig.2.- Iron abundance distributions in units of the solar abundance as a function of the distance from the cluster center. (a) model B. (b) model C.

Fig.3.- Iron abundance distributions in units of the solar abundance as a function of the distance from the cluster center for model D.

Fig.4.- Iron abundance distributions in units of the solar abundance as a function of the distance from the cluster center for model E.

Fig.5.- Iron abundance distributions in units of the solar abundance as a function of the distance from the cluster center. (a) model F. (b) model G.

Fig.6.- Iron abundance distributions in units of the solar abundance as a function of the distance from the cluster center. (a) Coma, A1060, Virgo and Centaurus. (b) A496, A2142, A2199 and AWM7. (c) A1795, MKW3s, Perseus and Hydra-A.

TABLE 1

SUMMARY OF MODEL PARAMETERS

| Model | $\dot{M}_c$ ($M_\odot$/yr) | $\beta$ | $r_{\rm st}$ (kpc) | $n_{\rm cD}(0)$ (cm$^{-3}$) | $R_{\rm eff}$ (kpc) | $t_{\rm d}$ ($10^8$yr) | $\tilde{u}_{RcD}$ | $\tilde{u}_a$ | $\tilde{Z}$ |
|---|---|---|---|---|---|---|---|---|---|
| A1-1 | 0 | 0.7 | 0 | ... | 0 | ... | ... | 0 | 1.4 |
| A1-2 | 100 | 0.7 | 0 | ... | 0 | ... | ... | 0.2 | 1.4 |
| A1-3 | 500 | 0.7 | 0 | ... | 0 | ... | ... | 0.8 | 1.3 |
| A1-4 | 1000 | 0.7 | 0 | ... | 0 | ... | ... | 1.7 | 1.2 |
| A2-1 | 0 | 0.5 | 0 | ... | 0 | ... | ... | 0 | 1.7 |
| A2-2 | 100 | 0.5 | 0 | ... | 0 | ... | ... | 0.1 | 1.6 |
| A2-3 | 500 | 0.5 | 0 | ... | 0 | ... | ... | 0.7 | 1.4 |
| A2-4 | 1000 | 0.5 | 0 | ... | 0 | ... | ... | 1.4 | 1.3 |
| B1 | 0 | 0.7 | 0 | ... | 0 | ... | ... | 0 | 1.9 |
| B2 | 100 | 0.7 | 0 | ... | 0 | ... | ... | 0.2 | 1.7 |
| B3 | 500 | 0.7 | 0 | ... | 0 | ... | ... | 0.8 | 1.4 |
| B4 | 1000 | 0.7 | 0 | ... | 0 | ... | ... | 1.7 | 1.3 |
| C1 | 0 | 0.7 | 30 | ... | 0 | ... | ... | 0 | 1.4 |
| C2 | 100 | 0.7 | 30 | ... | 0 | ... | ... | 0.2 | 1.5 |
| C3 | 500 | 0.7 | 30 | ... | 0 | ... | ... | 0.8 | 1.8 |
| C4 | 1000 | 0.7 | 30 | ... | 0 | ... | ... | 1.7 | 2.2 |
| D1 | 0 | 0.7 | 0 | 180 | 0 | ... | 0 | 0 | 5.1 |
| D2 | 100 | 0.7 | 0 | 180 | 0 | ... | 0.5 | 0.2 | 5.3 |
| D3 | 500 | 0.7 | 0 | 180 | 0 | ... | 2.4 | 0.8 | 1.3 |
| D4 | 1000 | 0.7 | 0 | 180 | 0 | ... | 4.7 | 1.7 | 1.2 |
| E1 | 100 | 0.7 | 0 | 90 | 0 | ... | 0.8 | 0.2 | 1.4 |
| E2 | 100 | 0.7 | 0 | 360 | 0 | ... | 0.3 | 0.2 | 5.5 |
| F1 | 0 | 0.7 | 0 | 180 | 5 | 100 | 0 | 0 | 5.1 |
| F2 | 0 | 0.7 | 0 | 180 | 10 | 100 | 0 | 0 | 2.1 |
| F3 | 0 | 0.7 | 0 | 180 | 30 | 100 | 0 | 0 | 1.0 |
| G1 | 0 | 0.7 | 0 | 180 | 5 | 1 | 0 | 0 | 5.1 |
| G2 | 0 | 0.7 | 0 | 180 | 10 | 1 | 0 | 0 | 5.1 |
| G3 | 0 | 0.7 | 0 | 180 | 30 | 1 | 0 | 0 | 3.6 |

TABLE 2

OBSERVATIONAL DATA

| Cluster | $z$ | $a$ (kpc) | $r_{\rm c}$ (kpc) | $n_\beta$ ($10^{-3}$cm$^{-3}$) | $n_{\rm d}^a$ ( $10^{-3}$cm$^{-3}$) | $\beta$ |
|---|---|---|---|---|---|---|
| Coma | 0.0232 | $310^f$ | $43^a$ | $3.0^f$ | 2.7 | $0.63^f$ |
| A1060 | 0.0114 | $100^g$ | $67^a$ | $4.5^g$ | 8.8 | $0.67^g$ |
| Virgo | 15Mpc | $7.1^j$ | $49^a$ | $43^b$ | 165 | $0.436^b$ |
| Centaurus | 0.0109 | $95^c$ | $79^a$ | $9^c$ | 22.1 | $0.45^c$ |
| A496 | 0.032 | $220^i$ | $177^a$ | $6.77^i$ | 9.8 | $(1)^i$ |
| A2142 | 0.0899 | $510^i$ | $204^a$ | $4.93^i$ | 8.5 | $(1)^i$ |
| A2199 | 0.0309 | $140^g$ | $181^a$ | $8.8^g$ | 14.1 | $0.68^g$ |
| AWM7 | 0.017 | $190^d$ | $100^e$ | ... | ... | ... |
| A1795 | 0.0616 | $300^g$ | $266^a$ | $5.8^g$ | 16.8 | $0.73^g$ |
| MKW3s | 0.0434 | $190^d$ | $120^e$ | ... | ... | ... |
| Perseus | 0.0184 | $280^g$ | $192^a$ | $4.5^g$ | 42 | $0.57^g$ |
| Hydra-A | 0.0522 | $145^h$ | $160^h$ | $6.5^h$ | 12.6 | $0.7^h$ |

| Cluster | $\dot{M}_{\rm c}$ ($M_\odot$/yr) | $N_{500}$ | $\sigma_r$ (km/s) | iron peak |
|---|---|---|---|---|
| Coma | $2^a$ | 28 | $1010^k$ | no |
| A1060 | $9^a$ | 11 | $608^k$ | no |
| Virgo | $10^a$ | 10 | $573^l$ | yes |
| Centaurus | $50^c$ | 13 | $586^m$ | yes |
| A496 | $121^a$ | 14 | $705^k$ | yes |
| A2142 | $244^a$ | 29 | $1241^n$ | yes |
| A2199 | $150^a$ | 18 | $795^o$ | yes |
| AWM7 | $(40)^e$ | 13 | $849^p$ | yes |
| A1795 | $512^a$ | 27 | $773^k$ | no |
| MKW3s | $(100)^e$ | ... | ... | no |
| Perseus | $393^a$ | 32 | $1277^k$ | no |
| Hydra-A | $600^h$ | ... | ... | no |

Note: In AWM7 and MKW3s $\dot{M}_{\rm c}$ is the mass deposition rate at the innermost observational bin. In A496 and A2142, $\beta$ is assumed to be 1. We used $H_0 = 50$km s$^{-1}$ Mpc$^{-1}$.

*a*. Edge & Stewart 1991a, *b*. Takano 1990, *c*. Matilski et al. 1985, *d*. Kriss

et al. 1983, *e*. Canizares et al. 1983, *f*. Hughes et al. 1988, *g*. Jones & Forman 1984, *h*. David et al. 1990, *i*. Abramopoulos & Ku 1983, *j*. Fabricant & Gorenstein 1983, *k*. Zabludoff et al. 1990, *l*. Binggeli et al. 1987, *m*. Lucey et al. 1986, *n*. Hintzen & Scot 1979, *o*. Gregory & Thompson 1984, *p*. Materne & Hopp 1983.

TABLE 3

$\tilde{u}_{RcD}$, $\tilde{u}_a$ AND $\tilde{Z}$ OF EACH CLUSTER

| Cluster | $\tilde{u}_{RcD}$ | $\tilde{u}_a$ | $\tilde{Z}$ |
|---|---|---|---|
| Coma | ... | 0.001 | 1.3 |
| A1060 | ... | 0.1 | 1.1 |
| Virgo | ... | 1.5 | (3.9) |
| Centaurus | 0.6 | 0.3 | 4.4 |
| A496 | 0.6 | 0.2 | 3.8 |
| A2142 | 1.0 | 0.03 | 1.1 |
| A2199 | 0.7 | 0.3 | 2.2 |
| AWM7 | 0.5 | 0.1 | 2.7 |
| A1795 | 1.4 | 0.2 | 1.2 |
| MKW3s | 2.5 | 0.3 | 1.0 |
| Perseus | 1.3 | 0.2 | 1.0 |
| Hydra-A | 3.2 | 1.7 | 1.3 |